\documentclass[times,twocolumn,final]{elsarticle}
\usepackage{setspace}

\graphicspath{{./}}

\usepackage{amsmath}
\usepackage{breqn}
\usepackage{soul,color}
\usepackage[dvipsnames]{xcolor}
\usepackage{pict2e}
\usepackage{array}
\usepackage[mathscr]{eucal}
\usepackage{upgreek}
\usepackage{latexsym}
\usepackage{siunitx}
\usepackage[version=4]{mhchem}
\usepackage{wasysym}
\usepackage{textcomp}
\usepackage[bottom]{footmisc}
\usepackage{caption}
\usepackage{tcolorbox}
\usepackage{blindtext}
\usepackage{hyperref}
\usepackage{graphicx}
\usepackage{subcaption}
\usepackage[export]{adjustbox}
\usepackage{wrapfig}
\usepackage{tabularx}
\usepackage{dblfloatfix}
\DeclareMathAlphabet{\mathpzc}{OT1}{pzc}{m}{it}

\usepackage{graphicx}
\usepackage{amssymb}

\usepackage{lineno}

\journal{Combustion and Flame}

\begin{document}
\begin{frontmatter}

\title{Stabilization of premixed NH$_3$/H$_2$/air flames via bluff-body flame holders}


\cortext[cor1]{Corresponding author: lukas.gaipl@ovgu.de}
\author[label1]{L. Gaipl\corref{cor1}}
\author[label1]{W. Guan}
\author[label2]{G. Guggilla}
\author[label2]{A. Kropman}
\author[label2]{F. Beyrau}
\author[label1]{D. Thévenin}

\address[label1]{Lab. of Fluid Dynamics and Technical Flows, University of Magdeburg "Otto von Guericke", Magdeburg 39106, Germany}
\address[label2]{Lab. of Technical Thermodynamics, University of Magdeburg "Otto von Guericke", Magdeburg 39106, Germany}

\date{11 February 2026}

\begin{abstract}
The stabilization mechanisms of fully premixed NH$_3$/H$_2$/air flames anchored near a bluff-body are investigated using experiments and direct numerical simulations. Particular attention is given to the interplay between preferential diffusion, heat release, flow recirculation, and turbulence–flame interaction. Comparison between non-reactive and reactive configurations shows that thermal expansion significantly alters the flow field, increasing the recirculation zone length by about 40\% and the shear layer width by roughly 50\% at the end of the recirculation region. Excellent agreement between measurements and simulations for mean and fluctuating axial velocities validates the numerical framework. Detailed analysis of the flame structure reveals a distinctive stabilization mechanism at the flame root: preferential hydrogen diffusion produces a localized diffusion flame branch that enhances radical production and strengthens anchoring. The chemical structure follows a sequential combustion process, with hydrogen primarily consumed in the shear layer, followed by ammonia cracking and the main heat release zone. Near the bluff-body, heat release is concentrated within the recirculation region, while downstream areas are increasingly affected by turbulence and velocity fluctuations. The roles of curvature and strain are quantified to assess stretch effects along the flame front. Convex curvature near the flame root promotes hydrogen enrichment and locally increases burning rates, reinforcing stabilization. In contrast, concave curvature and elevated stretch downstream of the recirculation zone weaken the flame and signal a transition toward a turbulence-dominated regime. Probability density functions of strain and curvature further distinguish dominant stretch mechanisms across flame regions. Overall, stabilization arises from coupled feedback between recirculation-induced heat exchange and rapid hydrogen oxidation, sustaining an intermediate ammonia reaction zone. These findings provide insight into carbon-free NH$_3$/H$_2$ flame stabilization and guidance for robust, low-emission combustion system design.
\end{abstract}


\end{frontmatter}



\section*{Novelty and significance statement}
\label{sec:NovSignState}
This work presents the first combined DNS and experimental investigation of a premixed NH$_3$/H$_2$/air flame stabilized behind a bluff-body. While bluff-body flame stabilization has been extensively studied for conventional fuels, the staged combustion behavior of ammonia–hydrogen mixtures, where ammonia must first crack to produce hydrogen, introduces fundamentally different stabilization mechanisms. Understanding these effects is crucial for developing robust and efficient combustors that can operate with carbon-free fuels across a wide range of conditions relevant to future applications. By resolving the detailed flame structure and reaction pathways, this study provides new insights into how fuel cracking and hydrogen release interact with recirculation and mixing processes, extending classical stabilization theory to carbon-free, staged combustion systems. Detailed analysis of turbulence-flame interaction reveals different mechanisms that govern flame stabilization at the flame root close to the bluff-body and at the end of the recirculation zone. 




\section{Introduction\label{sec:introduction}} \addvspace{10pt}
Flame stabilization remains a fundamental challenge in propulsion and power-generation systems, as it governs both combustion efficiency and operational safety~\cite{shanbhogue_lean_2009}. Various stabilization strategies have been developed for premixed combustion, among which bluff-body flame holders play a central role. By inducing a large recirculation zone (RZ) downstream of the obstacle, they provide a continuous source of hot products and radicals that promote flame anchoring under a wide range of operating conditions~\cite{lovett_review_2011}. Such configurations are particularly effective in high-velocity applications, including gas turbines and ramjets, where stable combustion must be maintained under lean and variable inflow conditions. For intrinsically slow-burning fuels such as NH$_3$/H$_2$ mixtures, which exhibit long chemical timescales and are therefore prone to blow-off~\cite{guan_revisiting_2024, guan_reduced_2025, awad_characteristics_2023}, the use of bluff-body stabilization represents a promising pathway towards high power-density, carbon-free combustion technologies.

Classical theories of flame stabilization via bluff-bodies are based on the interplay of two key processes: (1) the chemical kinetics controlling ignition and extinction, and (2) the fluid mechanics governing transport and mixing~\cite{williams_flame_1948, longwell_flame_1953, winterfeld_processes_1965}. These frameworks often rely on simplified energy or mass balance models, in which stabilization is interpreted as the result of interactions between the recirculation zone and the incoming reactant stream through the shear layer in the wake of the bluff-body. The resulting correlations, often expressed in terms of Damköhler numbers, relate the RZ length ($L_{RZ}$) or residence time in the shear layer to the characteristic ignition delay time of the fuel–air mixture~\cite{shanbhogue_lean_2009, zukoski_experiments_1956}. Two stages were characterized close to blow-off~\cite{nair_near-blowoff_2007}: 1) the offset of flame blow-off occurred with local extinction of the flame sheet. The extinguished flame sheets propagate along the flame, until they reignite to a continuous flame surface. In stage 2) of the blow-off, the flame front oscillated in an asymmetric mode. For reactive flow with large heat release across the shear layer, vorticity was attenuated due to dilatation resulting from large temperature gradients. It was shown that in the event of a blow-off the flame is partially quenched by excessive stretch in the shear layer, enabling fresh gases to enter the RZ~\cite{chaudhuri_blowoff_2010}. The flame sheet along the shear layer extinguishes and reactions take place in the RZ until the shear layer does not reignite anymore. Experimental studies of methane flames~\cite{kariuki_measurements_2012, dawson_visualization_2011} revealed structural changes close to blow-off, where a transition from a "V-shape" towards an "M-shape" of the flame sheet occurred due to entrainment of cold reactants into the recirculation zone. 

Preferential diffusion plays a central role in turbulent flame stabilization, with markedly different behaviors depending on the mixture Lewis number. For mixtures with Le $\leq1$, preferential diffusion generally enhances local burning rates and promotes flame propagation, while for Le $\geq1$ it tends to weaken flame propagation and increase susceptibility to quenching~\cite{poinsot_theoretical_2001}. The interaction between strain and Lewis number has been identified as a key factor controlling blow-off mechanisms~\cite{vance_effect_2019}. For Le $\leq1$ flames, extinction is most often associated with high stretch or strain rates that overcome the enhanced diffusive-thermal feedback. For Le $\geq1$ flames, blow-off has frequently been linked to local mixture enleanment and increased heat losses, although the relative importance of these processes is still debated. Several studies on Le $\geq1$ flames~\cite{chaudhuri_blowoff_2010, roy_chowdhury_effects_2018} also report that excessive strain can induce local extinction, enabling cold reactants to penetrate the recirculation zone and further weaken stabilization. Under such conditions, both flame speed and flame temperature decrease with increasing strain~\cite{law_combustion_2006}.

For NH$_3$/H$_2$ fueled flames, preferential diffusion effects are particularly important because of the strong diffusivity contrast between hydrogen and ammonia. Experiments have reported slightly wider lean blow-off limits than for methane or propane flames, despite lower global flame speeds and flame temperatures, which has been attributed to hydrogen preferential diffusion~\cite{su_lean_2024, su_behaviour_2024}. Fuel consumption is also spatially segregated: hydrogen is mainly consumed in a thin shear-layer region near the bluff-body, whereas ammonia reacts over a broader zone along the flame surface. Owing to its high diffusivity and burning velocity, hydrogen can sustain combustion at the flame root even close to blow-off. Ammonia-containing mixtures additionally exhibit a transition from V-shaped to M-shaped topologies and are often characterized by positive flame curvature, both of which contributing to extended stability limits. In contrast to hydrocarbon flames, which typically show a strong increase in strain rate near lean blow-off, ammonia flames do not display a similarly pronounced rise. Although bluff-body stabilization has been studied extensively for hydrocarbon fuels~\cite{kariuki_measurements_2012, longwell_flame_1948, williams_properties_1953, penner_recent_1957, wright_bluff-body_1959, maxworthy_mechanism_1962, fetting_turbulent_1958, fujii_comparison_1981, giacomazzi_coupling_2004}, systematic investigations for NH$_3$/H$_2$ mixtures remain limited. Yet stabilization is expected to be critical for these fuels because their ignition and burning characteristics are highly temperature-sensitive~\cite{awad_characteristics_2023}. Hydrogen enrichment not only widens lean blow-off limits and increases burning rates compared to methane~\cite{wiseman_comparison_2021}, but also reduces ignition delay times and enhances overall reactivity~\cite{li_numerical_2017}. Similar stabilization trends have been observed in premixed swirled combustors operating on NH$_3$/H$_2$/air blends~\cite{khateeb_stability_2020, khateeb_stability_2021}. These features motivate dedicated studies of NH$_3$/H$_2$/air flames stabilized by bluff-bodies to clarify their chemical structure, stabilization mechanisms, and turbulence–flame interactions.

 The present study employs experiments as well as direct numerical simulations (DNS) to investigate the combustion dynamics of premixed NH$_3$/H$_2$/air flames stabilized by bluff-bodies. Emphasis is placed on identifying the distinct features of premixed NH$_3$/H$_2$/air combustion and elucidating the mechanisms of flame stabilization. The findings provide new physical insights into the turbulence-chemistry interactions governing flame anchoring in low-reactivity fuels, offering guidance for the design of stable, efficient, and carbon-free combustion systems. 

The paper is structured as follows: the studied configuration as well as experimental methods are detailed in Sec.~\ref{sec:Config}, while numerical approaches are summarized thereafter (Sec.~\ref{sec:NumSetup}). The flow field of DNS is compared with experimental measurements to gain insights into the differences of non-reactive and reactive flow and further validate the DNS data (Sec.~\ref{subsec:FlowfieldAnalysis}). Finally, the flame structure is investigated using the validated DNS in Sec.~\ref{subsec:FlameStruct}. Implications of preferential diffusion of hydrogen resulting in a locally separated consumption of ammonia and hydrogen with consequences for flame stabilization are analyzed, before drawing conclusions in Sec.~\ref{sec:Conlusion}.

\section{Configuration and Experimental Setup\label{sec:Config}} 
\begin{figure}[ht!]
    \centering
    \includegraphics[width=0.8\linewidth]{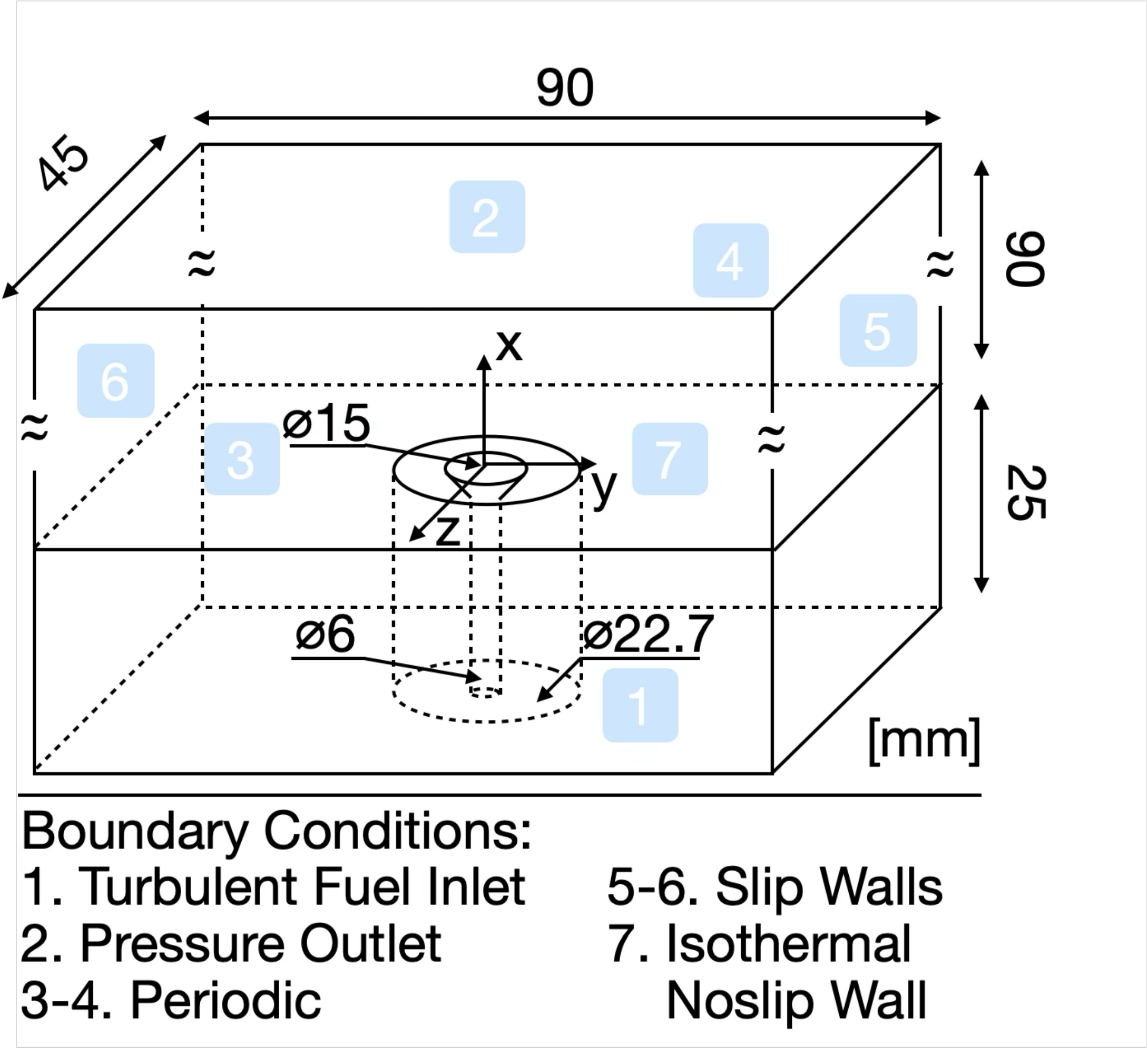}
    \caption{\footnotesize Geometry and boundary conditions.}
    \label{fig:ExpConfig}
\end{figure}

\addvspace{10pt}
The experimental setup consists of a burner capable of stabilizing a premixed turbulent flame on a bluff-body with a diameter of $D_{BB}=15$~mm and a burner throat diameter of 22.7~mm. The geometry and boundary conditions are shown in Fig.~\ref{fig:ExpConfig}. The annular velocity, $U=4.31$~m/s, is determined by dividing the volumetric flow rate by the annulus duct area, resulting in Re$=\rho U D_{BB} / \mu= 4688 $. 

\begin{figure*}[!b]
    \centering
    \includegraphics[width=0.75\linewidth]{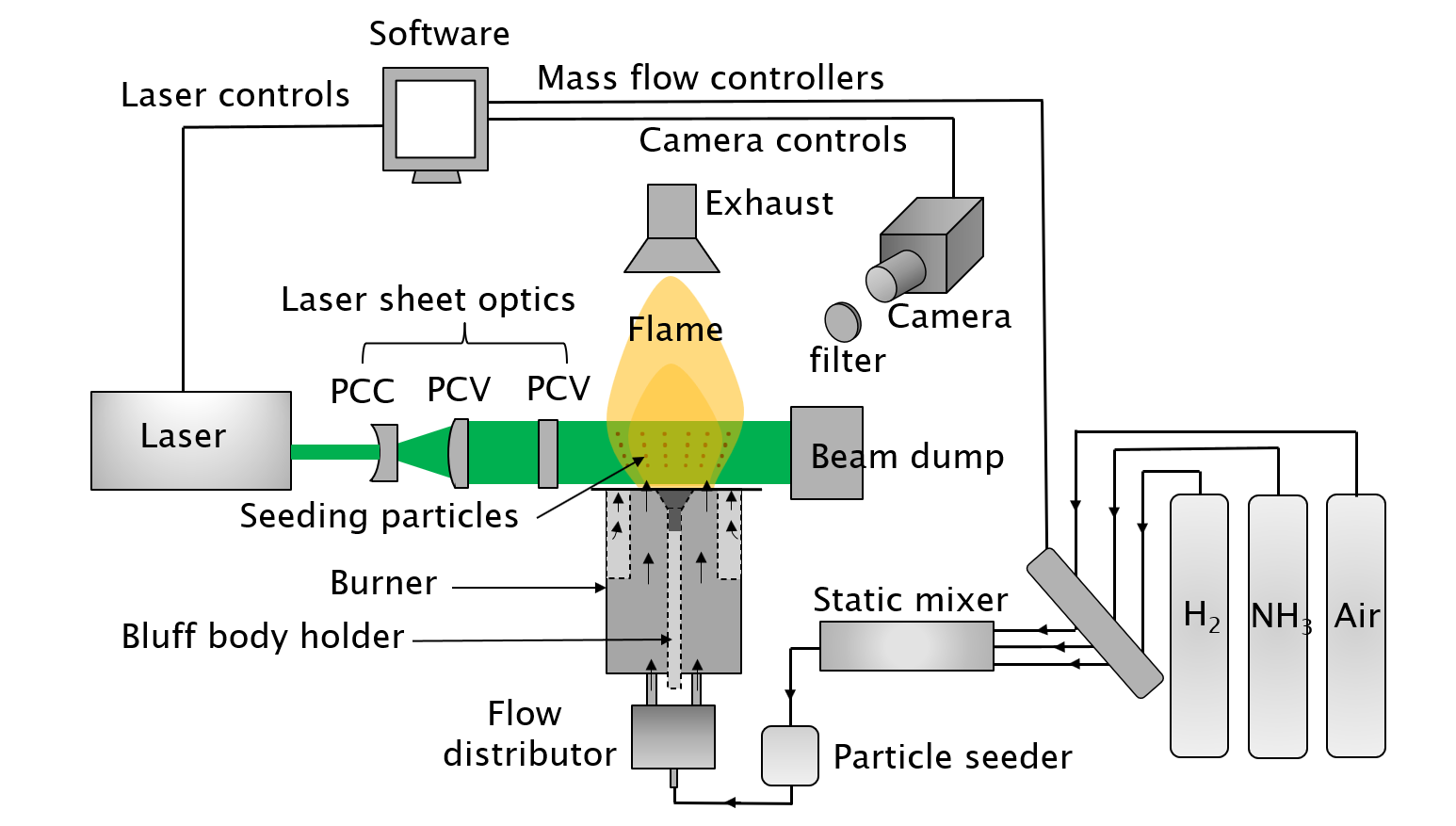}
    \caption{\footnotesize Experimental setup and measurement arrangement.}
    \label{fig:ExpSetup}
\end{figure*}
High-speed Particle Image Velocimetry (PIV) is employed to capture the time-resolved flow field under both cold-flow conditions and during NH$_3$/H$_2$/Air combustion. The experimental arrangement is presented in Fig.~\ref{fig:ExpSetup}. The particle image velocimetry (PIV) system operates at a repetition rate of 6~kHz in a double-pulsed configuration using a 532~nm laser. At this rate, each acquisition cycle consists of a double-frame image pair generated by two successive laser pulses separated by an inter-pulse delay of 80~\(\mu\)s. A laser sheet of 60~mm height and 0.8~mm measured thickness is generated using appropriate laser optics and aligned along the burner axis. Titanium dioxide (TiO$_2$) particles were used as seeding material for flow visualization, and images were acquired at a spatial resolution of 0.125~mm/pixel. The average particle diameter in the clusters was 250~nm (LaVision, Germany), corresponding to a Stokes number less than $1 \times 10^{-3}$, indicating that the particles closely follow the flow. The bluff body, fabricated from an Inconel alloy by means of additive manufacturing to allow for subsequent geometric modifications, is mounted on an adjustable holder that establishes an annular flow of the fuel–air mixture. A thermocouple is inserted through a drilled hole in the holder and positioned 1.5~mm below the bluff-body surface to monitor the near-surface temperature. This measurement provides the required boundary condition for the numerical simulations.

The quantities of ammonia, hydrogen, and air were controlled via a LabVIEW interface, with mass flow controllers (EL-FLOW series, Bronkhorst, The Netherlands) ensuring precise regulation of the flow rates. The controllers have a manufacturer-specified accuracy of $\pm0.5\%$ of reading plus $\pm0.1\%$ of full scale (SEMI E56) and were calibrated at reference conditions of $20\,^{\circ}\mathrm{C}$ and $1\,\mathrm{atm}$. The premixed fuel–air mixture is homogenized in an in-house designed static mixer and subsequently supplied to the burner via a flow distributor. The distributor separates the flow into four parallel channels, ensuring a uniform delivery of the mixture to the burner plenum. Ignition is initiated using an external pilot burner, which is removed after flame stabilisation. PIV system operation and image post-processing are performed using Davis 8.4 software (LaVision, Germany). The estimated Karlovitz, Dahmköhler and integral-scale Reynolds numbers from DNS at injector outlet indicate a premixed flame in the thin reaction zones regime with relatively low turbulence. It is therefore expected that hydrogen diffusion and curvature dominate strain effects. 

For the velocity analysis, 2000 double frames recorded at 6\,kHz were used after confirming statistical independence with respect to the sample size. This corresponds to a total acquisition time of $0.32\,\mathrm{s}$. Background noise and reflections were removed by subtracting a moving average image. Velocity fields were computed using adaptive multipass cross-correlation. The interrogation window was progressively reduced from $64 \times 64$ pixels over two passes with 75\% overlap to $16 \times 16$ pixels for non-reactive flow and $24 \times 24$ pixels for reactive flow over four passes with 50\% overlap. This procedure yielded final spatial resolutions of 1.0~mm for non-reactive flow and 1.5~mm for reactive flow. Erroneous vectors were identified and replaced using a standard median filter. The flow rate uncertainty was estimated from three repeated measurements. The calculated mean flow rate was $59.17\,\mathrm{LPM}$ (litres per minute) with a standard deviation of $\pm 0.76\,\mathrm{LPM}$ ($1\sigma$), corresponding to a relative uncertainty of approximately $1.3\%$. This relative uncertainty was propagated to the equivalence ratio and inlet velocity, resulting in an uncertainty of $\pm 0.01$ for $\phi = 0.8$ and approximately $\pm 1.3\%$ for the annular velocity. Flow conditions for the studied setup are summarized in Table~\ref{tab:Conditions}.

\begin{table}[ht!] \footnotesize
\caption{Key parameters and conditions for experiments and DNS.}
\centerline{\begin{tabular}{lc}
\hline 
Description (units)         & Value      \\
\hline
Annular velocity $U$ (m/s)      & 4.31          \\
Reynolds number (-)         & 4688 \\
Integral-scale Reynolds number (-) & 3.7\\
Karlovitz number (-) & 5.7 \\
Damköhler number (-) & 0.3 \\
Equivalence ratio $\phi$ (-)         & 0.8 \\
H$_2$ volume fraction $X$ (-)         & 0.2 \\
Injection temperature $T_\infty$ & 300 (K)  \\
Injection pressure $p_\infty$ & 101325 (Pa)  \\
\hline 
\label{tab:Conditions}
\end{tabular}}
\end{table}

\section{Numerical Setup\label{sec:NumSetup}} \addvspace{10pt}
Simulations are performed with the low-Mach DNS code DINO~\cite{abdelsamie_towards_2016}, that solves the Poisson equation using a Fast-Fourier-Transformation approach. A sixth-order finite-difference method is used for spatial discretization and a third-order explicit Runge–Kutta method is employed for temporal integration. 
The computational domain is uniformly discretized by $450\times1153\times865$ grid points, resulting in an isotropic resolution of 100~$\mu$m. The Kolmogorov scale $\eta$ = $(\nu^3/\epsilon)^{1/4}$, with $\nu$ the kinematic viscosity and $\epsilon$ the energy dissipation rate, was evaluated locally and 95\% of all grid cells fulfill the DNS criterion $\Delta x<2\cdot \eta$~\cite{pope_turbulent_2000}. This resolution leads also to at least 14 grid points across the flame front. A proper representation of the boundary layer at the lip of the bluff-body is ensured by an in-house high-order immersed boundary method (IBM)~\cite{chi_directional_2020, ou_directional_2022}. All embedded walls are described as well with IBM. The bluff-body surface is treated as an isothermal wall with $T=600$~K, corresponding to experimental measurements with a thermocouple located 1.5 mm beneath the bluff-body surface.
Stability of the simulations is controlled by setting the CFL and Fourier numbers to CFL=0.15 and Fo=0.1, resulting in a timestep of $\Delta t=3\cdot10^{-7}$ s.
NH$_3$/H$_2$/air chemistry is modeled using the analytically reduced kinetics from Guan et al.~\cite{guan_reduced_2025}. The mechanism is derived from the detailed NUIG-2023 mechanism~\cite{zhu_combustion_2024} and consists of 17 species, 10 quasi-steady state species and 180 reactions. It provides high accuracy across a wide range of conditions while showing a significant speedup compared to detailed chemistry. 
Species diffusion is accounted for using the Hirschfelder-Curtiss approximation while the Soret and Dufour effects are neglected, as curvature and strain were found to dominate local flame speed variation and emissions~\cite{chi_effects_2023}.
DNS has been carried out during 15 flow-through times in total. The last 5 flow-through times were used to compute time-averaged quantities that are analyzed in what follows.

\section{Results\label{sec:Results}} \addvspace{10pt}
Two complementary aspects are investigated in this work. First, an overview of the flow features emphasizing the coupling between the recirculation region and approaching gases is proposed and a detailed comparison of experiments and simulations is carried out. Non-reactive and reactive cases are analyzed to evaluate local differences in the flow-field that impact flame stabilization. The chemical flame structure is then described by examining key radicals to delineate the main zones of mixing, ammonia cracking and combustion. The leading edge of the flame is further analyzed to understand the stabilization mechanisms that are driven by preferential diffusion of hydrogen and ammonia cracking together with strain and curvature effects.
\\
\subsection{Flow Field Analysis\label{subsec:FlowfieldAnalysis}} 
\begin{figure}[ht!]
    \centering
    \includegraphics[width=1.0\linewidth]{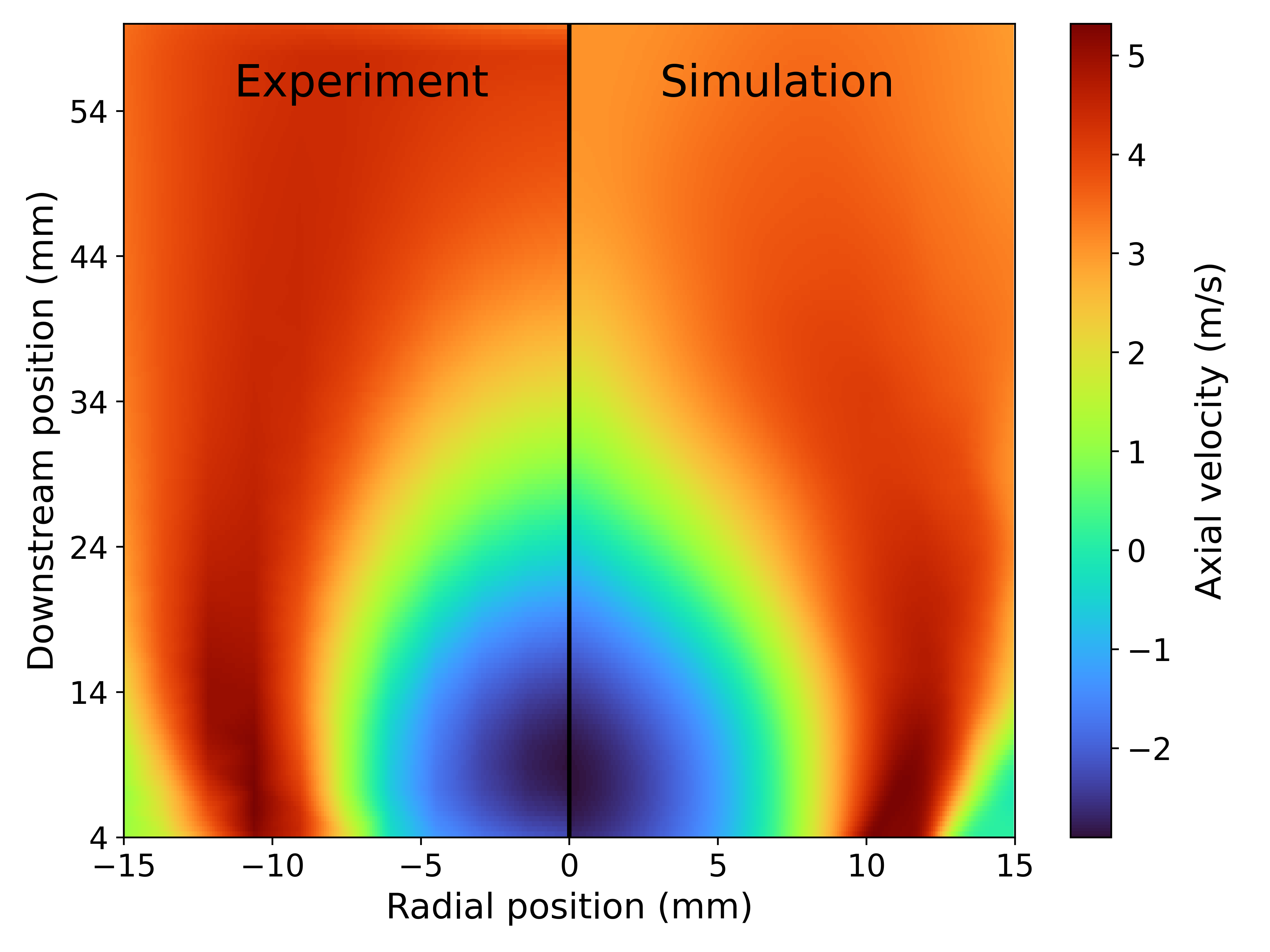}
    \caption{Comparison of experimental (left) and numerical (right) axial velocity fields for the reactive case.}
    \label{fig:VelFieldComparison}
\end{figure}

\addvspace{10pt}
The bluff-body flow field is analyzed to assess the predictive capability of the DNS and to clarify the hydrodynamic mechanisms that govern flame stabilization. Figure~\ref{fig:VelFieldComparison} presents the mean axial velocity for the reactive case. Imaging starts from 4 mm above the bluff-body to avoid optical reflections in the PIV measurements, with axial position $x=0$ corresponding to the flat surface of the bluff body.

\begin{figure}[ht!]
    \centering
    \includegraphics[width=0.9\linewidth]{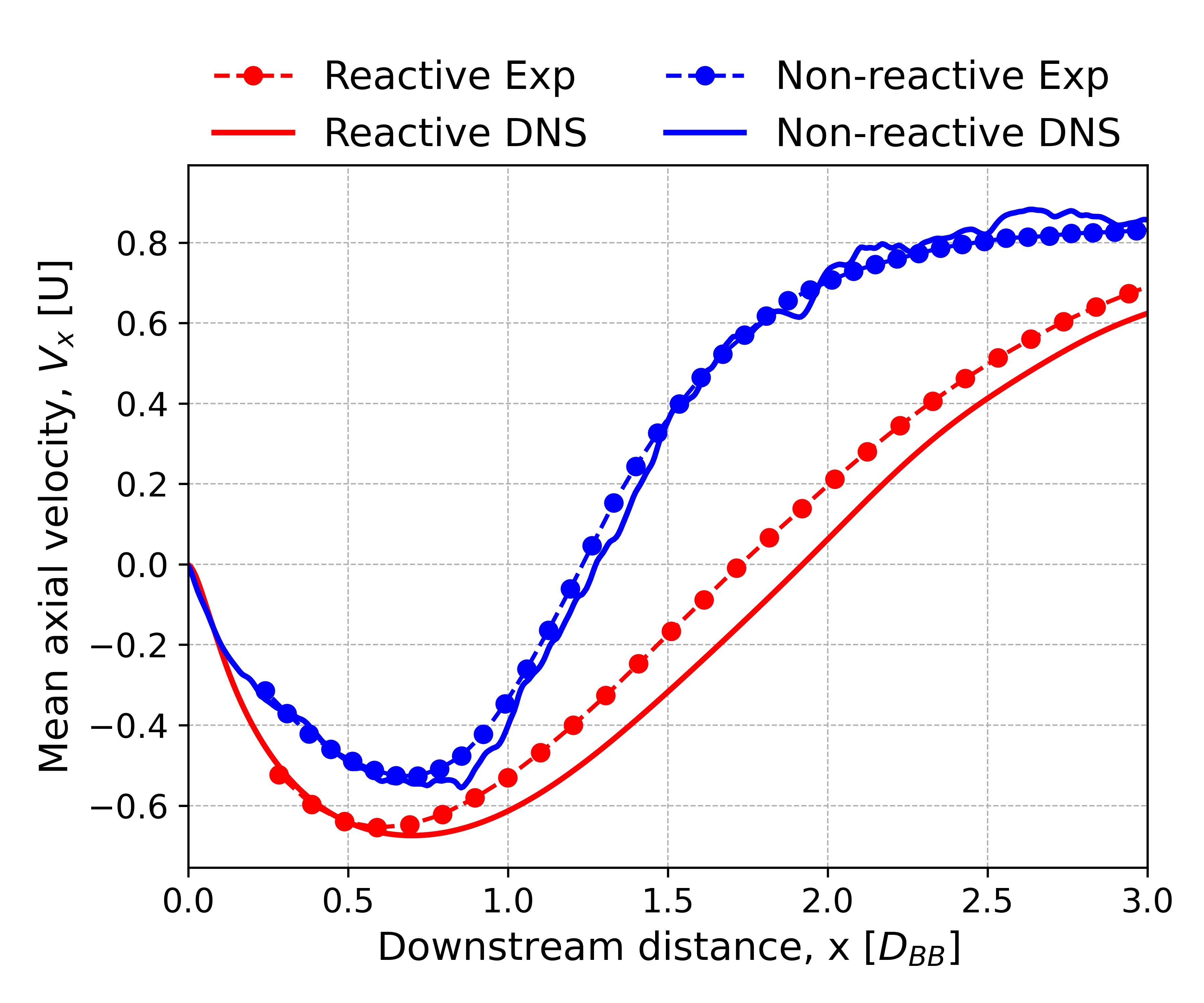}
    \caption{\footnotesize Axial velocity normalized by annular velocity (Tab.~\ref{tab:Conditions}) on a central line through the bluff-body for experiments and simulations. Axial velocity is normalized by the annular velocity ($U=4.31$~m/s) and downstream distance by the bluff-body diameter ($D_{BB}=15$~mm).}
    \label{fig:Recirculation_zone_length}
\end{figure}

The recirculation zone length is a global parameter that determines flame stability, as shown in classical theories~\cite{williams_flame_1948, longwell_flame_1953, winterfeld_processes_1965}. Its length is evaluated on the centerline of the bluff-body, starting from the bluff-body surface ($x=0$) to the point where axial velocity is zero (Fig.~\ref{fig:Recirculation_zone_length}). The non-reactive case shows excellent agreement between experiments and simulations and the recirculation zone length coincides at $x=1.2 D_{BB}$. For the reactive setup slight differences are observed, displaying $L_{RZ} = 1.75 D_{BB}$ for the experiments and $L_{RZ} = 1.9 D_{BB}$ for the simulations, respectively. As thermal boundary conditions strongly impact the flame behaviour, the used approach of an isothermal bluff-body alters the results and might be insufficient. Spatial variations of wall temperatures are highly possible due to the complex surrounding flow, leading to differences between experiments and simulations in recirculation zone length. To get a better agreement, a proper thermal description of the bluff-body is necessary. For this purpose, a conjugate heat transfer model has been implemented and validated in the code~\cite{guan_ghost-cell_2025}. Corresponding simulations will reveal the impact of realistic thermal boundary conditions on flame topology and properties. 

Acceleration of the flow through the annulus produces a high-momentum jet that separates at the bluff-body lip, forming a turbulent shear layer that encloses the inner recirculation zone. The DNS reproduces the experimental hydrodynamic features with close agreement. Both experiment and simulation show axial velocities up to 4.31~m/s at the burner exit, indicating consistent annular acceleration and comparable bulk momentum flux. Discrepancies are primarily confined to the downstream end of the recirculation zone, where steeper velocity gradients are predicted by the DNS. Nevertheless, the recirculation-zone length (see later Fig.~\ref{fig:Recirculation_zone_length}) and radial extent (approximately 7.5~mm from the centerline) are well captured. The correct prediction of these large-scale structures indicates that the DNS resolves the dominant flow features controlling scalar transport, flame topology, and global stabilization behavior in this bluff-body configuration. 

\begin{figure*}[ht!]
    \centering
    \includegraphics[width=1.0\textwidth]{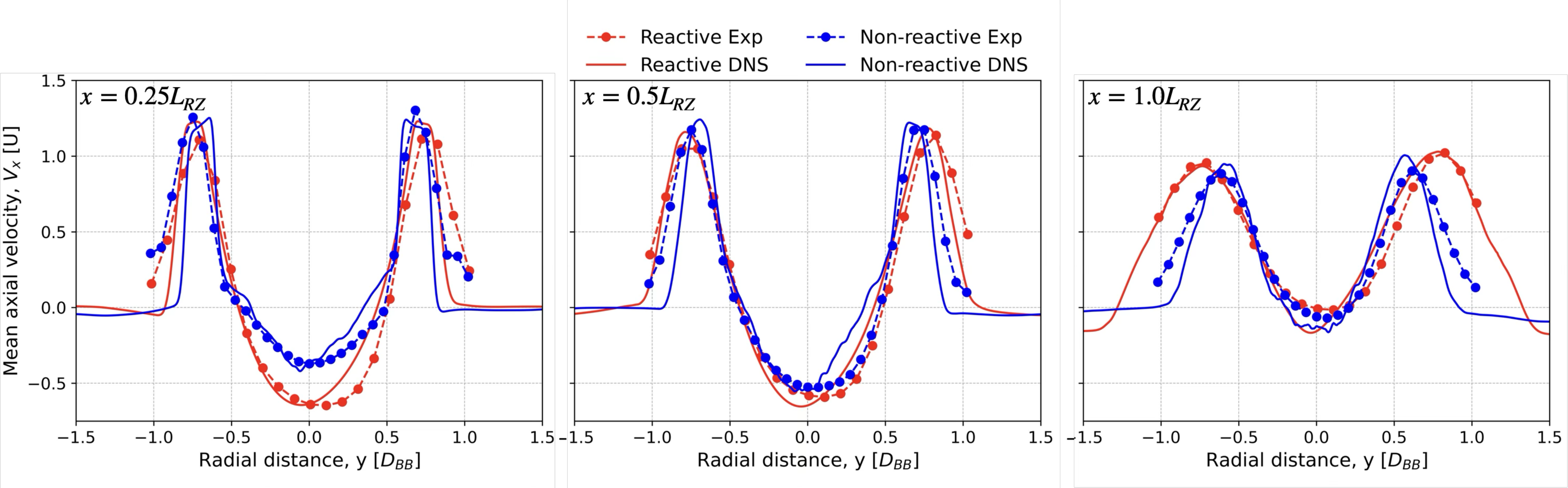}
    \caption{\footnotesize Axial mean velocity comparison for increasing downstream distances from left to right, normalized by the experimental recirculation zone length ($L_{RZ}$). Velocities are normalized by the annular velocity $U$(Tab.~\ref{tab:Conditions}).}
    \label{fig:MeanVelComp}
\end{figure*}

\begin{figure*}[ht!]
    \centering
    \includegraphics[width=1.0\textwidth]{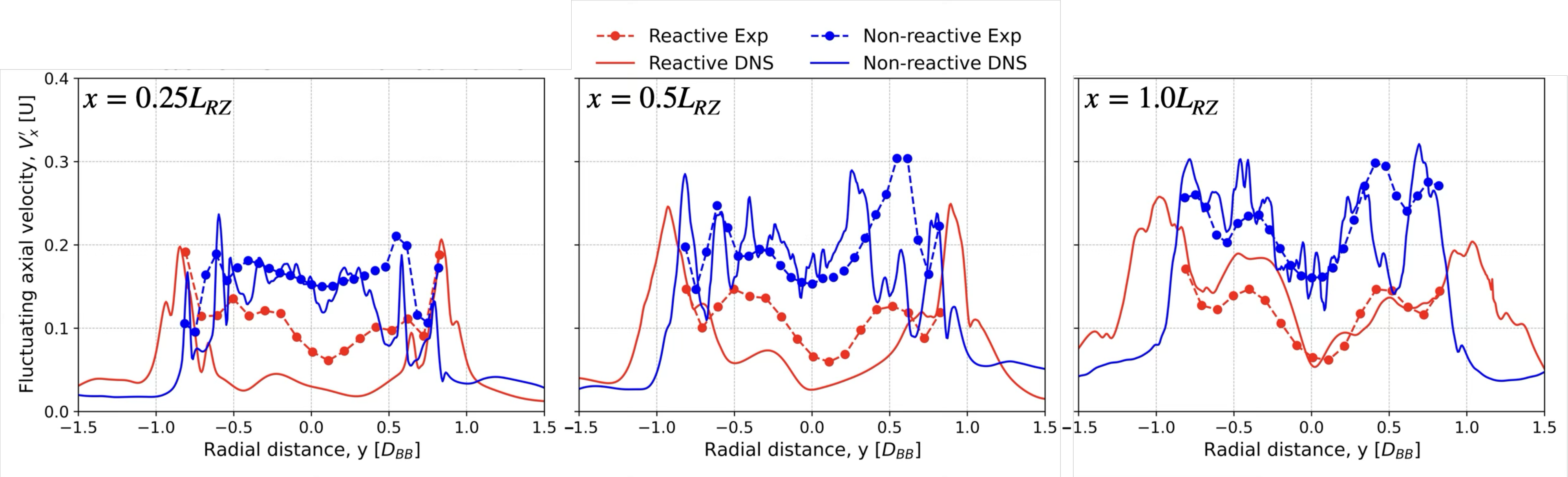}
    \caption{\footnotesize Axial velocity fluctuation comparison for increasing downstream distances from left to right, normalized by the experimental recirculation zone length ($L_{RZ}$). Velocities are normalized by the annular velocity (Tab.~\ref{tab:Conditions}).}
    \label{fig:StdVelComp}
\end{figure*}
The averaged axial velocity is further analyzed at downstream locations of $x = 0.25L_{RZ}$, $0.5L_{RZ}$, and $1.0L_{RZ}$ (Fig.~\ref{fig:MeanVelComp}), to assess the turbulent exchange in the shear layer, which governs the entrainment of hot burnt gases into the incoming fresh mixture. Capturing this exchange accurately is of utmost importance, as it directly controls the heat and mass transfer processes that determine flame stabilization and overall combustor performance. Experiments and simulations show close agreement for non-reactive and reactive cases. With increasing downstream distance, the peak in velocity flattens and the shear layer grows in radial direction. Clear differences are observed between non-reactive and reactive flow: while the mean axial velocity peaks at $1.25$ times the annular velocity ($U=4.31$~m/s) for both cases, the width of the shear layer is altered, growing twice as wide for the reactive case at the end of the recirculation zone. 

The turbulent exchange of heat and mass within the recirculation zone and shear layer is governed by the turbulent fluctuations, displayed in Fig.~\ref{fig:StdVelComp}. Peaks of axial velocity fluctuations are located at a similar location than the peak of mean axial velocity (Fig.~\ref{fig:MeanVelComp}), where the turbulent shear layer is dominant. A clear reduction of velocity fluctuations is observed for the reactive case, where flame dilatation leads to a turbulence reduction in the flow field. Overall, velocity fluctuations in the recirculation zone are decreased by 50~\%, while the shear layer is widened by around 50~\% compared to the non-reactive case, thus leading to a broader exchange of heat and mass at the interface of burnt and fresh gases. The measurements provide physically sound values of velocity fluctuations up to a radial extend of $0.8D_{BB}$, where the seeding particle density is high enough. 


\subsection{Flame Structure Analysis\label{subsec:FlameStruct}} 
\begin{figure}[h!]
    \centering
    \includegraphics[width=1.0\linewidth]{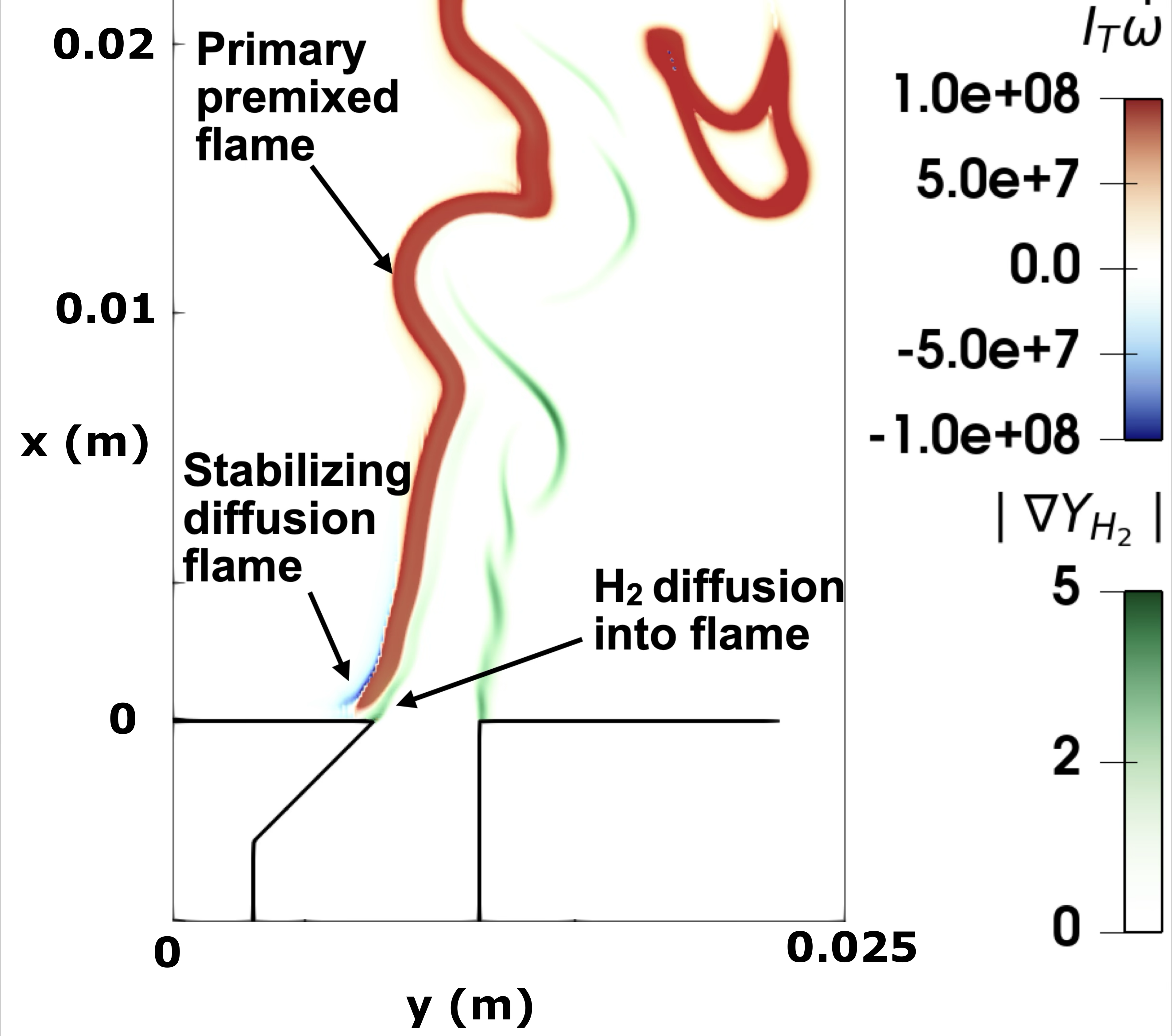}
    \caption{\footnotesize Zoom on flame anchoring displaying flame structure and hydrogen diffusion.}
    \label{fig:FlameStructure}
\end{figure}

\addvspace{10pt}
The global flame structure is analyzed in Fig.~\ref{fig:FlameStructure}, where local heat release rate is multiplied with the Takeno index~\cite{yamashita_numerical_1996} to visualize the combustion regime~\cite{gaipl_combustion_2025}:
\begin{equation}
    I_T = \frac{\nabla Y_F \cdot \nabla Y_{O_2}}{|\nabla Y_F \cdot \nabla Y_{O_2}|} 
\end{equation}
where $Y_F=Y_{NH_3}+Y_{H_2}$ is the fuel mixture mass fraction and $Y_{O_2}$ the oxygen mass fraction. The results reveal a predominantly premixed flame branch established in the shear layer between hot recirculating combustion products and the fresh mixture stream. A distinctive feature is observed at the root of the flame: large gradients of the hydrogen mass fraction are induced by preferential diffusion at the edge of the bluff-body, leading to a small diffusion flame branch. This feature is observed throughout the whole duration of the DNS analysis (5 flow-through times). 

\begin{figure}[ht!]
    \centering
    \includegraphics[width=1.0\linewidth]{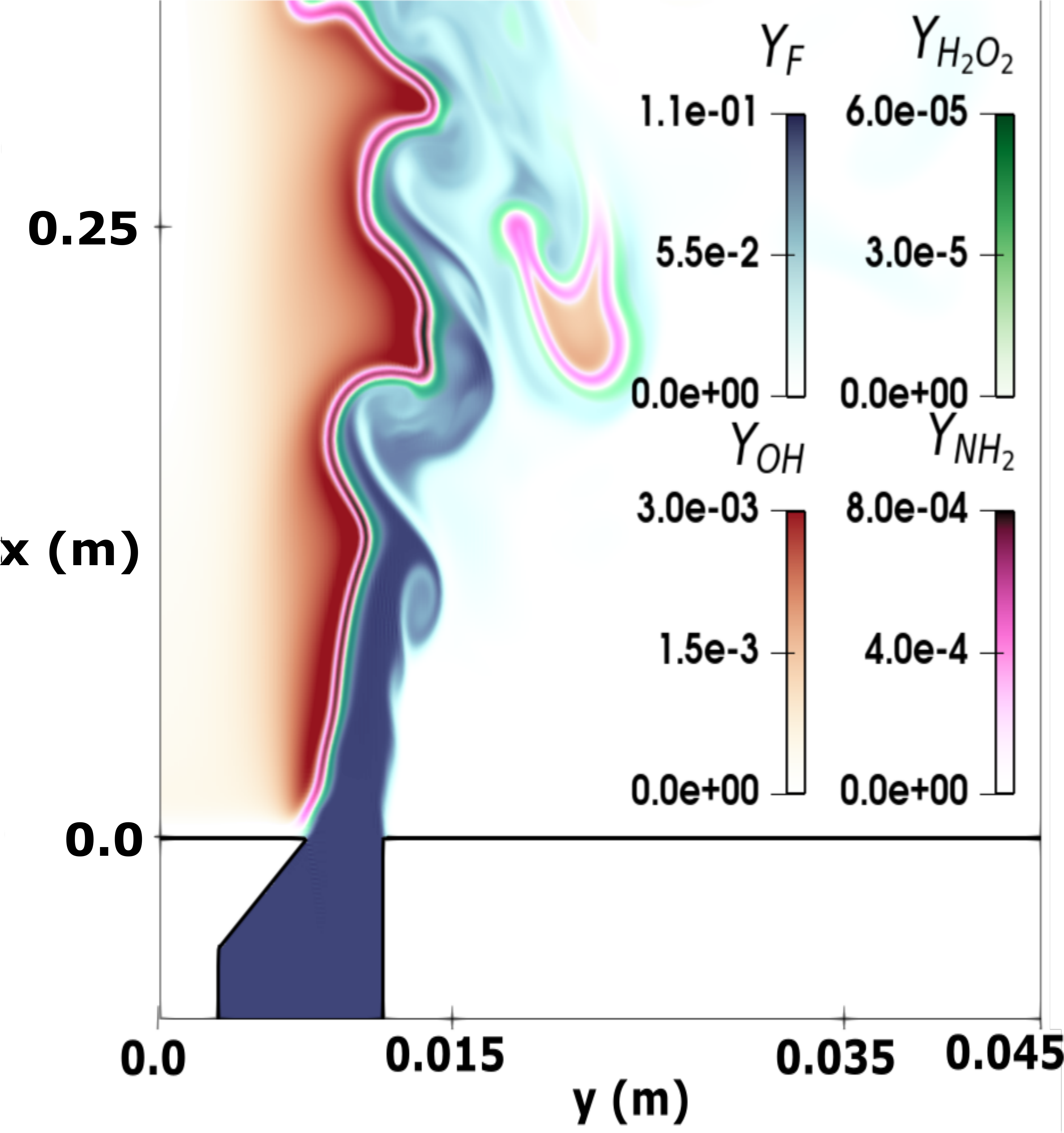}
    \caption{\footnotesize Instantaneous distribution of selected radical species and fuel mass fraction.}
    \label{fig:ChemFlameStructure}
\end{figure}

To investigate the flame structure obtained numerically, key intermediate radicals representative of different stages of the combustion process are further examined in a midplane cut through the domain (Fig.~\ref{fig:ChemFlameStructure}). A layered combustion process is observed: The hot recirculation zone transfers heat to the incoming cold gases through the shear layer, promoting ammonia decomposition in high-temperature regions, as indicated by elevated NH$_2$ mass fractions. The resulting hydrogen enriches the surrounding mixture, where it is consumed in a primary reaction zone that is oriented towards the center of the bluff-body (increased OH mass fraction). A secondary reaction zone occurs, as displayed by an increased H$_2$O$_2$ concentration near the outer edge of the recirculation region, where the hydrogen issued from ammonia is consumed. Local enrichment of the mixture by preferential diffusion of hydrogen (as shown in Fig.~\ref{fig:FlameStructure}) leads to enhanced radical production by local ammonia cracking. Resulting radicals consume excess oxygen near the recirculation zone and thereby support flame stabilization.

\begin{figure*}[ht!]
    \centering
    \includegraphics[width=1.0\textwidth]{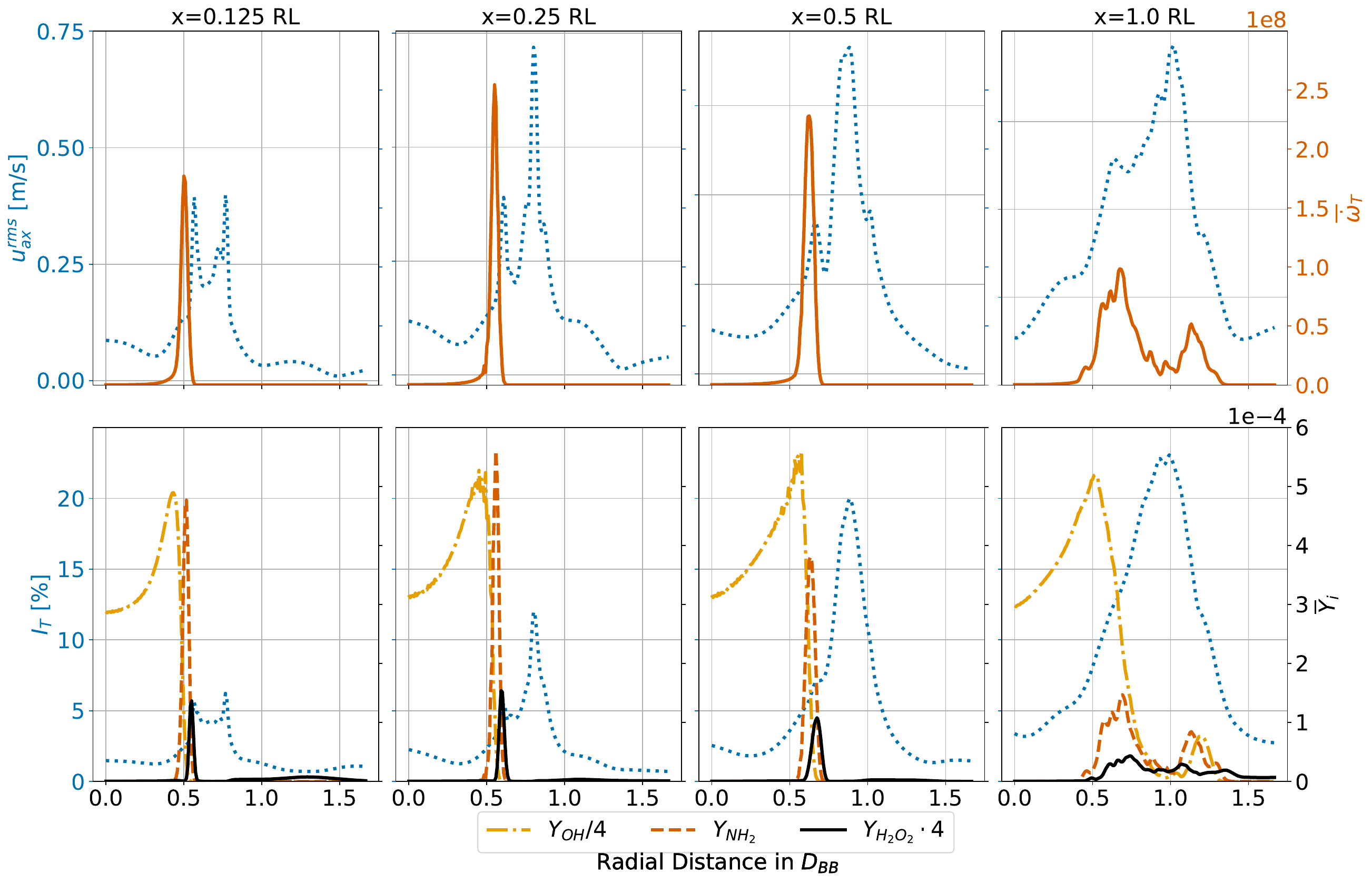}
    \caption{Analysis of the turbulent quantities interacting with the flame for increasing distance from bluff body from left to right. Top row: position of the mean heat release rate ($\overline{\dot{\omega}}_T$) with respect to axial velocity fluctuations ($u_{ax}^{rms}$). Bottom row: average chemical flame structure ($\overline{Y}_i$ for OH, NH$_2$, H$_2$O$_2$) with respect to turbulence intensity ($I_T$).}
    \label{fig:Hot_Flame_Analysis}
\end{figure*}
An in-depth analysis of the chemical flame structure is performed at increasing downstream distances along a radial line extending from the bluff-body center to a distance of $1.7D_{BB}$, reaching up to the center of the recirculation zone. Along these lines, turbulence statistics, heat release rate, and key chemical species are examined after building corresponding time averages of the mass fraction fields (Fig.~\ref{fig:Hot_Flame_Analysis}). Close to the bluff-body, the peak of heat release rate is located on the inner side of the shear layer, as evidenced by comparison with the peaks in axial velocity fluctuations. Hydrogen consumption, traced by elevated $Y_{H_2O_2}$, reaches a maximum within the shear layer at a radial distance of approximately $0.5D_{BB}$. Ammonia cracking follows further inside the recirculation zone, where increased temperatures, resulting from hot recirculated products and hydrogen reactions on the opposite side, promote its conversion. Further downstream, the heat release rate peak progressively shifts toward the inner part of the shear layer, aligning with the inner peak of axial velocity fluctuations. Near the end of the recirculation zone, enhanced shear-layer mixing distributes both ammonia and hydrogen consumption over a broader region, leading to a widened flame structure with two heat release rate peaks located on the inner and outer sides of the shear layer.

These observations point to two main mechanisms governing flame stabilization. First, at the flame root, the exchange between hot recirculated gases and the incoming fresh mixture is critical. Hydrogen, which can react at comparatively low temperatures, contributes to preheating the reactants and thus facilitates subsequent ammonia combustion. The combined effect of heat feedback from the recirculation zone and hydrogen oxidation stabilizes an ammonia reaction zone in between. In this region, hydrogen consumption is strongly modulated by turbulent shear, whereas ammonia consumption occurs closer to the maximum heat release rate zone and is therefore less exposed to intense stretch (Fig.~\ref{fig:Hot_Flame_Analysis}). Second, further downstream, the heat release rate peak aligns with local maxima in velocity fluctuations, indicating a regime dominated by turbulence–flame interaction. Here, turbulence-induced stretch can become excessive and promote local extinction. This behavior is consistent with experimentally observed flame kernels~\cite{chaudhuri_blowoff_2010}, which may intermittently reignite the shear layer prior to global blow-off.

 To evaluate the local strain and curvature, the flame front is defined using a temperature-based progress variable~\cite{yang_direct_2022}:
 \begin{equation}
     c_T = \frac{T-T_u}{T_{b}-T_u}\text{.}
 \end{equation}
 To trace the reactive part of the flame front, values are conditioned to $0.4<c_T<0.6$. 
 Curvature is evaluated in 2D or 3D according to~\cite{poinsot_theoretical_2001}:
\begin{align}
    K_c= -\nabla \vec{n} &= \left. -\left(\frac{\partial n_x}{\partial x} + \frac{\partial n_y}{\partial y}\right) \right| _{2D} \\
    & = \left. -\left(\frac{\partial n_x}{\partial x} + \frac{\partial n_y}{\partial y} + \frac{\partial n_z}{\partial z}\right) \right|_{3D}
\end{align}
and the tangential strain as:
\begin{equation}
    a_t = \nabla_t\cdot \vec{n} = (\delta_{ij}-n_in_j)\frac{\partial n_i}{\partial x_j}\text{.}
\end{equation}

\begin{figure}[ht!]
    \centering
    \includegraphics[width=0.5\textwidth]{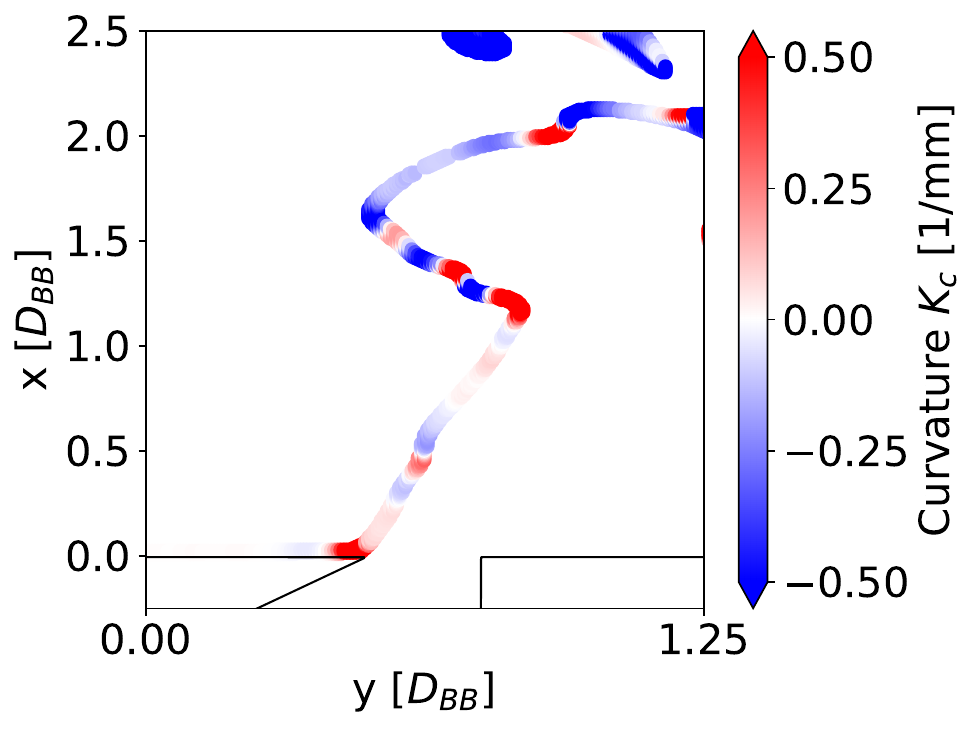}
    \caption{\footnotesize Flame front curvature close to the bluff-body in a midplane cut through the domain.}
    \label{fig:curvature}
\end{figure}

An instantaneous snapshot of flame front curvature (2D) is displayed in Fig.~\ref{fig:curvature} on a midplane cut through the domain. At the leading edge, positive curvature leads to increased hydrogen diffusion that subsequently stabilizes the flame further in this area. Downstream, larger curvature is observed overall and major parts of the flame front exhibit concave shapes, that are prone to a decrease in flame speed, being thus more sensitive to extinction via excessive flame stretch. Similar effects have been observed for NH$_3$/H$_2$/N$_2$ mixtures~\cite{su_behaviour_2024}, where convex curvature at the leading edge enhanced blow-off resistance compared to methane mixtures with equal flame speeds.

\begin{figure}[ht!]
    \centering
    \includegraphics[width=0.5\textwidth]{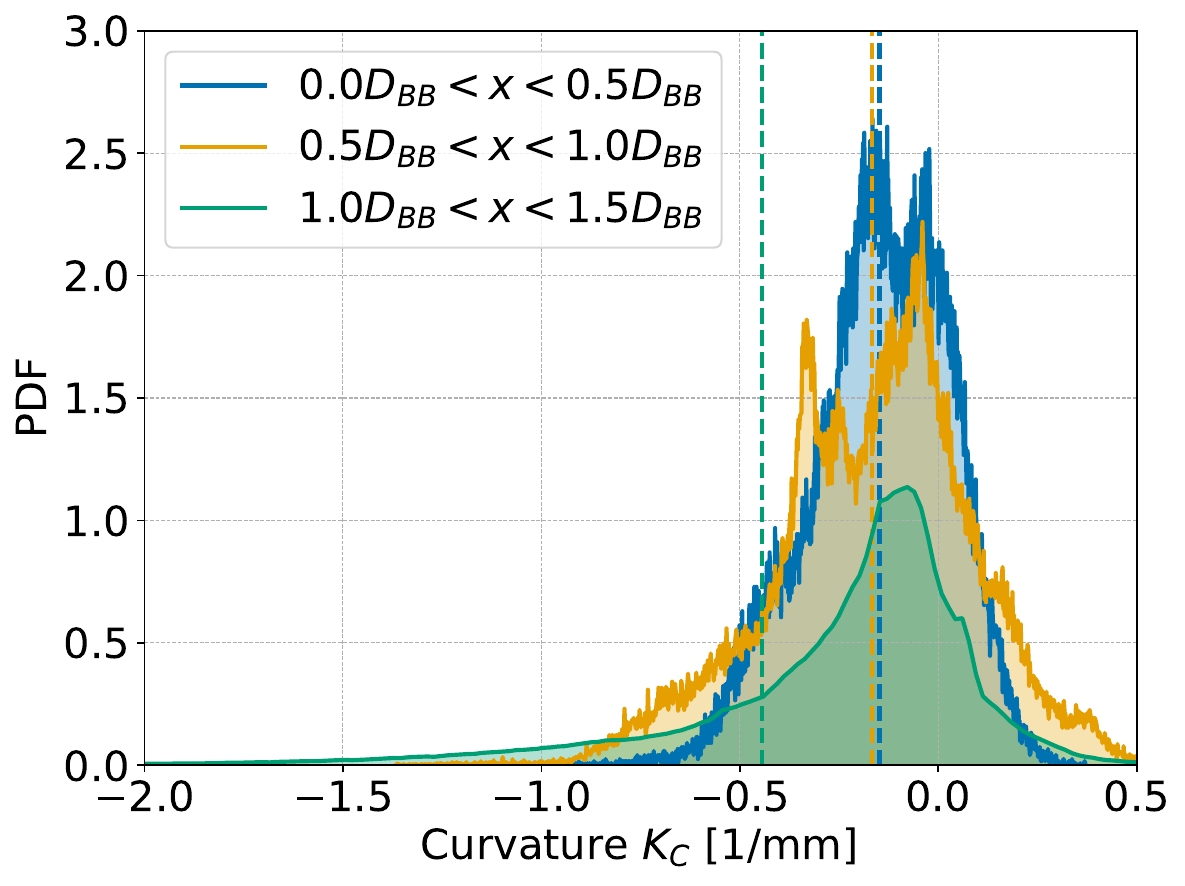}
    \caption{\footnotesize Flame front curvature pdfs in different regions of the flame. Dashed lines show the mean value.}
    \label{fig:pdf_curvature}
\end{figure}

Different regions of the flame are further investigated using the probability density function (pdf) of 3D-curvature (Fig.~\ref{fig:pdf_curvature}) and clear differences are observed. Close to the bluff-body ($x<0.5D_{BB}$), curvature exhibits mainly slightly negative values in a relatively narrow range ($|K_C|\leq 0.75$~1/mm). Further downstream, the pdfs broaden and larger positive and (even more so) negative curvature is observed. The pdfs thus support the findings of Fig.~\ref{fig:ChemFlameStructure} by a second argument: at the flame root, flame curvature exhibits smaller convex values that aid to flame stabilization, while close to the end of the recirculation zone, larger curvature is observed that destabilizes the flame front by excessive stretch and decrease in hydrogen diffusion (for concave values).

\begin{figure}[ht!]
    \centering
    \includegraphics[width=0.5\textwidth]{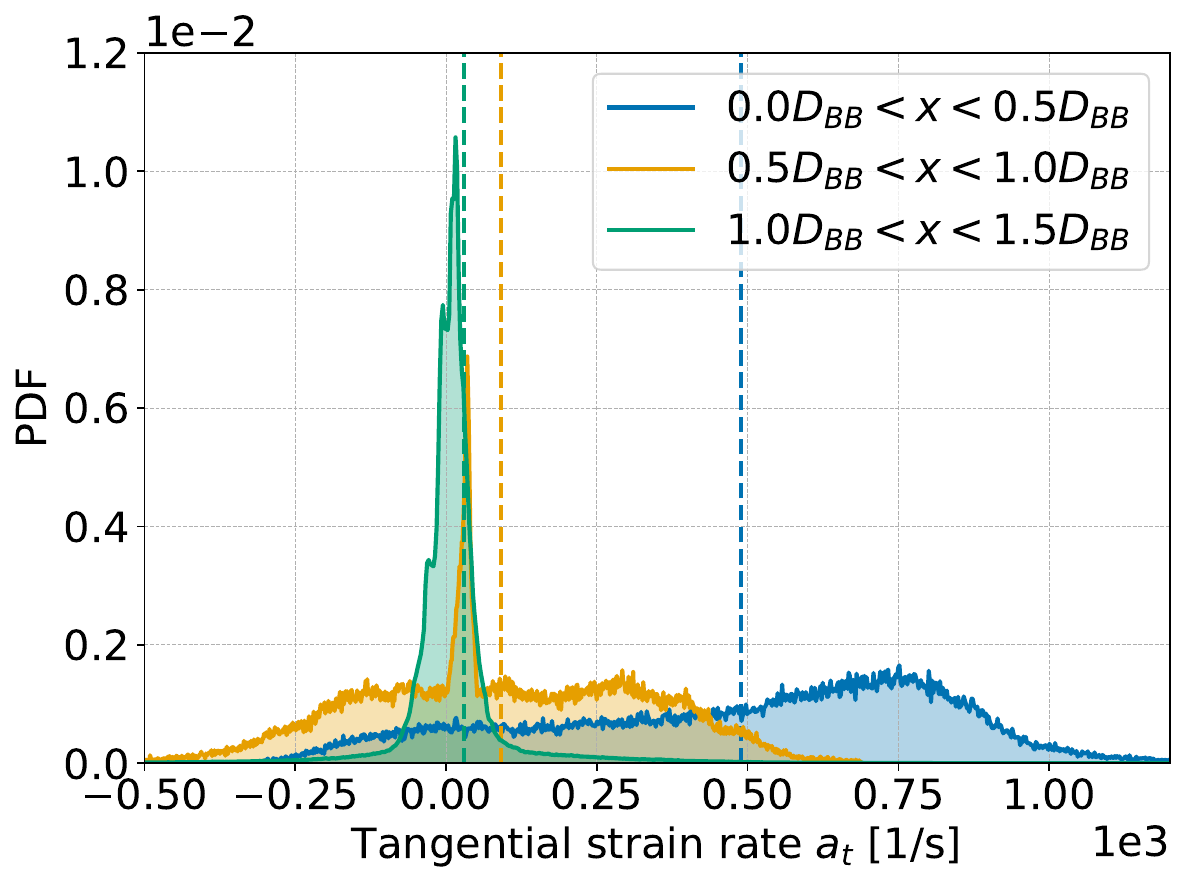}
    \caption{\footnotesize Pdfs of tangential strain rate in different regions of the flame. Dashed lines show the mean value.}
    \label{fig:pdf_strain}
\end{figure}

The contribution of tangential strain rate to stretch is evaluated in Fig.~\ref{fig:pdf_strain}. Close to the bluff-body, the shear layer separates and velocities exhibit highest values with sharp gradients (Fig.~\ref{fig:VelFieldComparison}) resulting in a widespread pdf of tangential strain rate up to $a_t>1000$~1/s. Further downstream, the tangential strain rate decreases and peaks around zero. Compared to curvature, that exhibits increased values in magnitude with downstream distance, the trend is inverted and highlights important aspects for the flame stabilization of bluff-body stabilized NH$_3$/H$_2$/air flames: the flame root is essentially controlled by the balance between large tangential flame strain and differential diffusion, while further downstream, the flame is subjected to higher values of flame front curvature. At the flame root, the flame front is wrinkled with a convex shape, adding to hydrogen diffusion into the flame front, thus stabilizing the flame. At the end of the recirculation zone ($L_{RZ}\approx 1.25$~$D_{BB}$), a lower tangential strain acts on the flame, while flame front curvature is enhanced by the developed shear-layer turbulence, and the flame gets locally quenched prior to global blow-off. 

\section{Conclusion \label{sec:Conlusion}} \addvspace{10pt}
Combined experimental measurements and direct numerical simulations were performed for a fully-premixed NH$_3$/H$_2$/air flame stabilized by a bluff-body in order to elucidate the mechanisms governing flame anchoring. A comparison of non-reactive and reactive configurations revealed pronounced modifications of the recirculation zone induced by thermal expansion. In the reactive case, flame dilatation increased the recirculation zone length by approximately 40~\%, accompanied by a radial widening of the shear layer by about 50~\%. Overall, the agreement between experiments and simulations was good for mean axial velocities and their fluctuations, thereby supporting the validity of the DNS framework and the imposed boundary conditions.

The flame structure was then analyzed in detail. A distinctive feature was identified at the flame root: a localized diffusion-flame branch originating from preferential hydrogen diffusion. This locally enriched mixture enhanced radical production and thereby promoted robust flame anchoring. Chemically, the flame exhibited a sequential structure. Hydrogen, stemming from ammonia cracking was consumed first within the shear layer, followed by ammonia cracking and the main heat release region, where H$_2$ was primarily consumed (characterized by elevated OH mass fractions). In the vicinity of the bluff-body, the peak of heat release rate was located on the inner side of the shear layer, inside the recirculation zone. Further downstream, the flame interacted with increasing velocity fluctuations and intensified turbulence levels.

The roles of curvature and strain were subsequently examined. Near the flame root, the flame was predominantly convex toward the reactants. This configuration enhanced preferential hydrogen diffusion into the reaction zone, increased the local burning velocity, and contributed further to flame stabilization. In contrast, near the end of the recirculation zone, the flame front exhibited mainly concave curvature, which promoted excessive stretch, reduced hydrogen enrichment, and locally weakened the flame. Probability density functions of strain and curvature clarified the spatial variation of stretch contributions. At the flame root, large tangential strain, associated with strong axial velocity gradients was dominant, whereas curvature effects prevailed near the downstream end of the recirculation zone.

These observations provide a coherent picture of bluff-body flame stabilization in premixed NH$_3$/H$_2$/air flames. Stabilization results from a combined feedback mechanism: heat exchange with the recirculation zone and rapid hydrogen oxidation sustain an intermediate ammonia reaction zone. Hydrogen consumption is strongly modulated by turbulent shear, while ammonia conversion occurs closer to the main heat release rate region and is therefore less exposed to intense stretch. At the end of the recirculation zone, peaks in heat release coincide with local maxima of velocity fluctuations, indicating a regime primarily governed by turbulence–flame interactions.

Future work will aim to deepen further the physical understanding and enhance the practical relevance of these findings. Advanced laser diagnostics will be employed to resolve local flame structure, temperature fields, and intermediate species. Complementary simulations including conjugate heat transfer modeling~\cite{guan_ghost-cell_2025} will assess flame anchoring for different high-temperature materials. In addition, emission characteristics, particularly NO$_x$ formation pathways will be investigated under confined and practically relevant operating conditions to quantify the environmental impact of NH$_3$/H$_2$ combustion. Finally, a systematic parametric study of bluff-body geometries will be conducted to generalize the identified stabilization mechanisms and support the design of robust, low-emission combustors.

\section*{CrediT Authorship Contribution Statement}
\label{sec:AuthContr}
\textbf{LG}: Conceptualization, Data Curation, Formal analysis, Writing - original draft
\textbf{WG}: Conceptualization, Data Curation, Formal analysis, Writing - review and editing
\textbf{GG}: Conceptualization, Data Curation, Formal analysis, Visualization, Writing - original draft
\textbf{AK}: Conceptualization, Formal analysis, Writing - review and editing
\textbf{FB}: Funding acquisition, Supervision, Writing - review and editing
\textbf{DT}: Funding acquisition, Supervision, Writing - review and editing

\section*{Declaration of Competing Interest}

The authors declare that they have no known competing financial interests or personal relationships that could have appeared to influence the work reported in this paper.

\section*{Acknowledgments}

Financial support of the Deutsche Forschungsgemeinschaft (DFG) within project SPP 2419 HyCAM (project number: 523880888) is greatly acknowledged. The computer resources provided by SuperMUC-NG project pn36gi at Leibniz Supercomputing Center Munich have been essential to obtain the results presented in this work.

\footnotesize
\baselineskip 9pt

\thispagestyle{empty}
\bibliographystyle{pci}
\bibliography{PCI_LaTeX}


\newpage

\small
\baselineskip 10pt


\end{document}